# A Spectrometric Approach to Measuring the Rayleigh Scattering Length for Liquid Scintillator Detectors

S.S. Gokhale,[a] R. Rosero,[a] R. Diaz Perez,[a] C. Camilo Reyes,[a] S. Hans,[a,b] and M. Yeh[a,c,*]

[a]*Chemistry Division, Brookhaven National Laboratory, Upton, New York 11973*
[b]*Chemistry Department, Bronx Community College, Bronx, New York 10453*
[c]*Instrumentation Division, Brookhaven National Laboratory, Upton, New York 11973*
*E-mail*: yeh@bnl.gov

ABSTRACT: Good optical transparency is a fundamental requirement of liquid scintillator (LS) detectors. Characterizing the transparency of a liquid scintillator to its own emitted light is a key parameter to determine the overall sensitivity of a large-volume detector. The attenuation of light in an optical-pure LS is dominated by Rayleigh scattering, which poses an intrinsic limit to the transparency of LS. This work presents a spectrometric approach of measuring the wavelength-dependent scattering length of liquids by applying the *Einstein-Smoluchowski* theory to a measurement of scattered light intensity. The scattering lengths of linear alkyl benzene (LAB) and EJ309-base (Di-isopropylnaphthalene, DIN) were measured and are reported in the wavelength range of 410 – 520 nm. The spectral peak of scintillation light emitted by a nominal LS is around 430 nm at which the scattering length for LAB and EJ-309-base was determined to be $27.9 \pm 2.3$ m and $6.1 \pm 0.6$ m respectively.

[*]Corresponding author.

# Contents

## 1. Introduction

Large-volume liquid scintillator (LS) detectors provide the necessary stopping power and sensitivity required for neutrino detection and have played a key role in the understanding of neutrino physics over the past decades [1, 2, 3, 4]. Typically, these large detectors have target masses ranging from tens of tons to kilotons. The corresponding volumes are such that scintillation light generated in the center of the detector may have to traverse several meters of the liquid medium before being collected by the photomultiplier tubes (PMTs) [5, 6]. Therefore, the transparency of the scintillation liquid to its own emitted light is imperative to determine the overall sensitivity of the detector. As the size of the detectors increases, the significance of optimizing the transmission of scintillation light through the liquid increases dramatically. The optical transparency can be optimized by ensuring high molecular purity of the solvent. However, there is a limit to the transparency that can be obtained under realistic conditions because of the Rayleigh scattering of light off the molecules of liquid.

    Linear alkyl benzene (LAB; $C_nH_{2n+1}$–$C_6H_5$, n = 10–13) has received much attention as a detection medium for nuclear and particle physics experiments. LAB has a relatively good light yield, with an emission peak at 284 nm that can be reemitted to longer wavelengths (~340nm) through multiple self-absorption and reemission processes. Its inherent properties of high flash point (130 °C), low toxicity, and long attenuation length favor LAB as a detection medium for large-volume scintillator detectors. The optical transmission and scintillation propagation of LAB have been studied extensively for its application in neutrino detections [7, 8, 9]; however, there has not been a systematic measurement of scattering length over the wide range of visible region.

    EJ309 (trade name under Eljen) is another scintillator material that is prepared from Di-isopropylnaphthalene (DIN, $C_{16}H_{20}$) based solvent and has gained great interest as a scintillation material due to its unique pulse-shape characteristics of the ionization particles for short baseline neutrino detectors [10]. It is a commercially available liquid scintillator, originally developed for neutron detection [11]. EJ309-base (EJ309, exclusive of fluor or wavelength shifter) is a nonflammable solvent with very low vapor pressure. It has a high light yield with an emission spectrum dominated by a peak at 360 nm after multiple self-absorption and reemission processes like LAB. Previous work has characterized the light collection and response of EJ309, however

most of these studies have been limited to smaller volumes [12, 13]. Figure 1 shows the emission spectra for both LAB and EJ309-base under UV excitation.

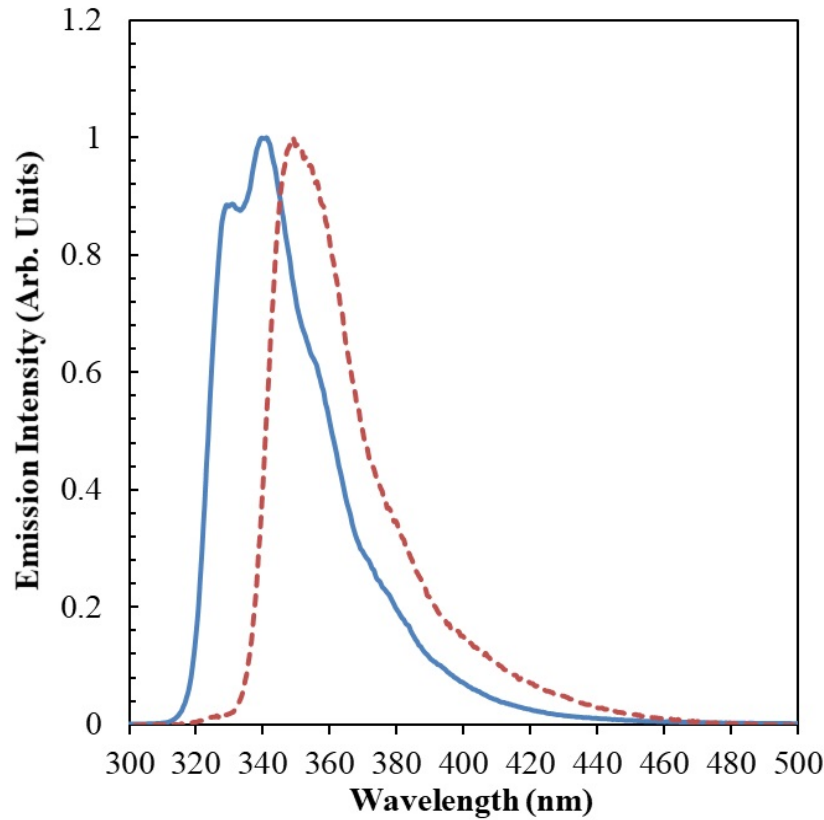

**Figure 1.** Emission spectra of pure LAB (solid) and EJ309-base (dash) with excitation λ = 260nm.

A nominal liquid scintillator (nLS) employed in most neutrino detectors is composed of three components: aromatic organic scintillator as the solvent, 2,5–diphenyloxazole (PPO) as the fluor, and 1,4–bis[2–methylstyryl] benzene (bis–MSB) as the wavelength shifter. The wavelength of the scintillation light emitted by a pure scintillator solvent is shifted to the visible region by PPO and bis-MSB. The spectral maximum of the scintillation light for an nLS is often controlled at peak, λ = 430 nm, which is within the range of wavelengths in which the quantum efficiency (QE) of the detector PMTs is most sensitive. To optimize the optical clarity in the 400 – 600 nm range, the scintillator solvent is subjected to multiple purification steps to remove both inorganic and organic impurities, which may present absorption bands in the wavelength range of interest. The self-absorption of the scintillation light by LAB or EJ309-base in this wavelength region is negligible, therefore the major component of the optical attenuation process is molecular scattering.

The attenuation of light in an nLS is composed of absorption and scattering [14]. A photon travelling through a liquid can either be absorbed or be deflected from its path when it interacts with the molecules of medium. The absorbed photons are either reemitted to different wavelengths or converted into heat and lost. Scattering on the other hand mainly changes the direction in which the photons are propagated, and the photons can eventually be collected by the surrounding PMTs. The charateric response of a large-volume liquid scintillator detector consequently depends on the precise knowledge of the optical attenuation property of the detection medium.

The attenuation length of a liquid can be described by the following formula:

$$\frac{1}{L} = \frac{1}{L_A} + \frac{1}{L_S} \tag{1.1}$$

where $L_A$ and $L_S$ are the absorption and scattering lengths, which are defined as the average distance that the probability of a photon traversing in a liquid medium without being absorbed or scattered is 1/e. It is difficult to directly measure the absorption length, but it can be determined by measuring the scattering and total attenuation lengths and substituting into equation 1.1. Previous experiments have estimated the Rayleigh scattering length and the total attenuation length of LAB by measuring the attenuation of light at limited wavelengths of interest only [9]. The experimental data for the Rayleigh scattering lengths crossing the full region of emitted wavelengths for both LAB and EJ309-base (or DIN) have not been reported.

The present work develops an approach to measuring the wavelength-dependent, scattering length for any ultraclean liquids by applying the *Einstein-Smoluchowski* theory to the scattered light intensity measured by a fluorescence spectrometer. This approach utilizes a spectrometric module, as opposed to a long-arm apparatus with pathlength of several meters, to measure the scattering length of scintillator solvents. The scattering lengths of LAB and EJ309-base are reported in the wavelength range of 410 – 520nm.

## 2. Theoretical Approach

The light scattering off the particles or molecules of a pure liquid predominantly undergoes Rayleigh scattering when the wavelength ($\lambda$) of the incident photons is much larger compared to all the dimensions of the scattering particle (or molecule). Therefore, the scattering of light off a volume of liquid can be expressed by the Rayleigh ratio, which is characterized as the volume scattering function $\beta(\theta)$ for a scattering angle of 90° [15]

$$R \equiv \beta(90°) = \left(\frac{I}{I_0}\right)\left(\frac{r^2}{V}\right) \tag{2.1}$$

where, $I$ is the intensity of the transversely scattered light, $I_o$ is the intensity of the incident light, $r$ is the distance from the scattering center to the point of observation, and $V$ is the scattering volume. The intensity of the scattered light can be divided into two components that are parallel and perpendicular to the scattering plane. The angular distribution of these components depends on the polarization of the scattered light induced by the scattering material, which is characterized by the depolarization ratio $\delta$ that allows for both isotropic and anisotropic contributions to the Rayleigh scattering amplitude.

In the case of an ideal isotropic liquid, the depolarization ratio is zero and the scattering length depends only on the Rayleigh ratio. In addition, if the liquid is perfectly uniform, the scattering will be eliminated by destructive interference except in the forward direction. However, because of the thermal motions of the molecules, the local density of a liquid is constantly fluctuating. The effect of these random density fluctuations is from the dielectric constant. As the refractive index is directly related to the dielectric constant, the fluctuations can then cause an increase in the scattering amplitude. To account for this effect of nonuniformity of the liquid on scattering, the isotropic Rayleigh ratio has been expressed by [16, 17, 18, 19, 20]

$$R_{iso} = \frac{\pi^2}{2\lambda^4}\left[\rho\left(\frac{\partial \varepsilon}{\partial \rho}\right)_T\right] kT\kappa_T \tag{2.2}$$

where, $\lambda$ is the wavelength of the scattered light, $\rho$ is the density of liquid, $\varepsilon$ is the average dielectric constant of the liquid, $k$ is the *Boltzmann constant*, $T$ is the temperature, and $\kappa_T$ is the isothermal compressibility. The total Rayleigh ratio ($R$) related to the isotropic Rayleigh ratio is further given by the *Cabannes* formula [21]

$$\frac{R}{R_{iso}} = \frac{6 + 6\delta}{6 - 7\delta} \tag{2.3}$$

The fluctuation on the amount of scattering in the dielectric constant can be evaluated using empirical relations between density and the refractive index ($n$). Several such relations have been published in the literatures [15, 22], but for organic liquids it was shown that the *Eykman* equation describes $\rho(\partial\varepsilon/\partial\rho)$ well [17, 22, 23, 24]. The *Eykman* formula is

$$\rho\left(\frac{\partial\varepsilon}{\partial\rho}\right)_T = \frac{(n^2 - 1)(2n^2 + 0.8n)}{n^2 + 0.8n + 1} \tag{2.4}$$

In a pure liquid medium, the attenuation of light due to scattering can be characterized in the terms of the Rayleigh scattering length ($l_{Ray}$), an equation for which was first proposed by Einstein based on the *Einstein-Smoluchowski* theory [18]. This equation for scattering length, which considers all the parameters which affect scattering as described above, is as follow

$$l_{Ray} = \left\{\frac{8\pi^3}{3\lambda^4}\left[\frac{(n^2 - 1)(2n^2 + 0.8n)}{n^2 + 0.8n + 1}\right]^2 kT\kappa_T \frac{6 + 6\delta}{6 - 7\delta}\right\}^{-1} \tag{2.5}$$

## 3. Optical Measurements

### 3.1 Rayleigh Scattered Light Intensity

The light scattering measurements were performed using a PTI QuantaMaster-8000 Fluorescence Spectrometer, which uses a 75 W xenon lamp as a light source. Figure 2 shows the schematic drawing of the instrument. The pure liquids used for the scattering length measurements were high purity (18.3 MΩ-cm) water purified using a Millipore Milli-Q water system, linear alkyl benzene (Cepsa, 500-Q) cleansed by thin-film temperature-dependent vacuum distillation, and EJ309-base pretreated by Eljen using ion-extraction columns.

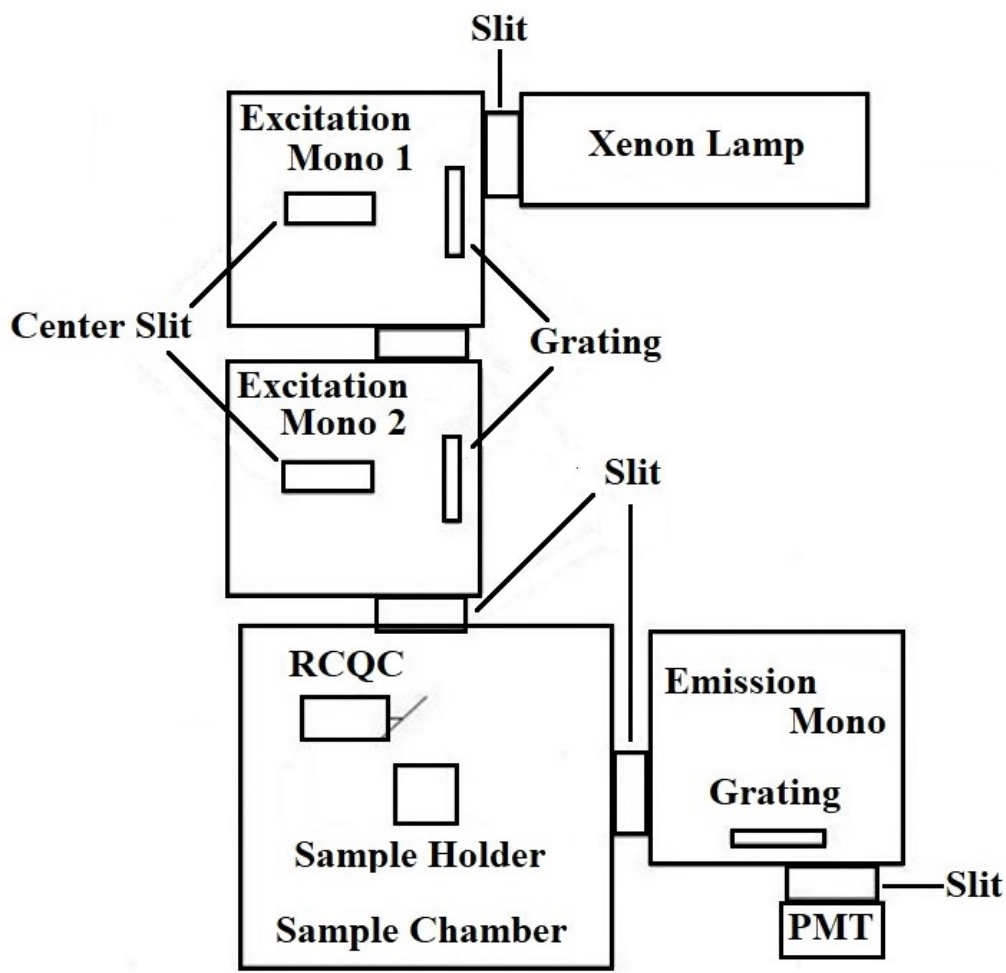

**Figure 2.** Schematic illustration of fluorescence spectrometer system used for measuring Rayleigh scattering where Mono – Monochromator, RCQC – Reference Channel Quantum Counter.

The incident light is scattered at a right angle by the liquid sample contained in a quartz cell. Suppression of stray light is a critical factor when accurately measuring the scattered light intensity. A combination of slits and lenses together reduce the divergence of the primary beam illuminating the liquid sample cell. The light wavelength of interest from the xenon lamp is selected by double additive, 350-mm focal length asymmetrical Czerny-Turner monochromators. The monochromators also ensure further stray light suppression with a rejection rate of $10^{-10}$. The intensity of the light emitted by the xenon lamp is not equal across the entire output spectrum. Therefore, the central part of the primary light beam from the monochromators passes an aperture and onto the sample, and the remainder is reflected towards the RCQC device, which has a reference diode detector. The reference diode detector provides a correction signal for any temporal fluctuation of the lamp intensity [25].

The spectrometer measures the intensity of the scattered light by means of a cooled photomultiplier tube (PMT). The scattered light is admitted to the PMT through an optical system consisting of lens, slit, and monochromator, which ensures a good delimitation of scattering volume and the solid angle of the cone of scattered light. The spectrometer applies a correction factor to compensate the variation in the detection efficiency of the optics, gratings, PMT, etc. for all the different light wavelengths. The scattered light intensity spectra were measured for each wavelength of incident monochromatic light. The intensity peak area corresponds to the total number of photons collected by the PMT over the acquisition period, while the peak height corresponds to the maximum intensity of the scattered light. Figure 3 shows the wavelength-dependent Rayleigh scattering responses for LAB.

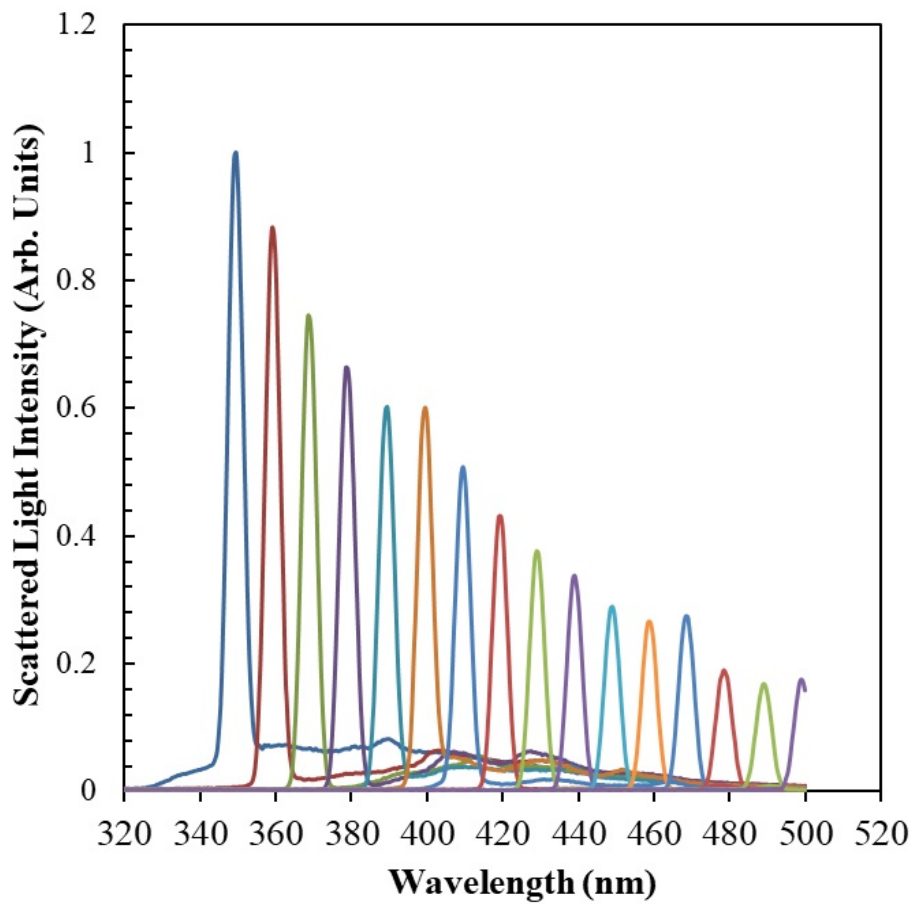

**Figure 3.** Scattered light intensity spectra for a LAB sample illuminated with different wavelengths of monochromatic light. The scattering intensity decreases as a function of wavelength by a theoretical factor of $\lambda^{-4}$.

The intensity of the Rayleigh scattered light given by the peak heights decreases with an increase in the wavelength and is roughly proportional to $\lambda^{-4}$. The Intensity peaks correspond to the intensity of scattered light for different wavelengths, however for wavelengths shorter than 440 nm the scattered light peaks appear to be mixed with the light emitted from the excitation-reabsorption-remission process.

Figure 4 shows that the effect of emitted light on the scattered light intensity in the region of 400–440 nm can be seen from the elevated baseline of longer wavelength tail of the Intensity, whereas the output spectra for incident light wavelengths >440 nm consist of only the Intensity distribution of the scattered light. Therefore, only wavelengths longer than 450 nm were used for the scattered intensity fitting to eliminate the interference of emitted light with the scattered length analysis for both LAB and EJ309-base.

The scattered light spectrum at each wavelength consist of pairs of peaks, the primary elastic Rayleigh scattering peak and a secondary inelastic Raman peak. The intensity of the Raman peak is a factor of ~30 smaller than that of the Rayleigh scattering peak.

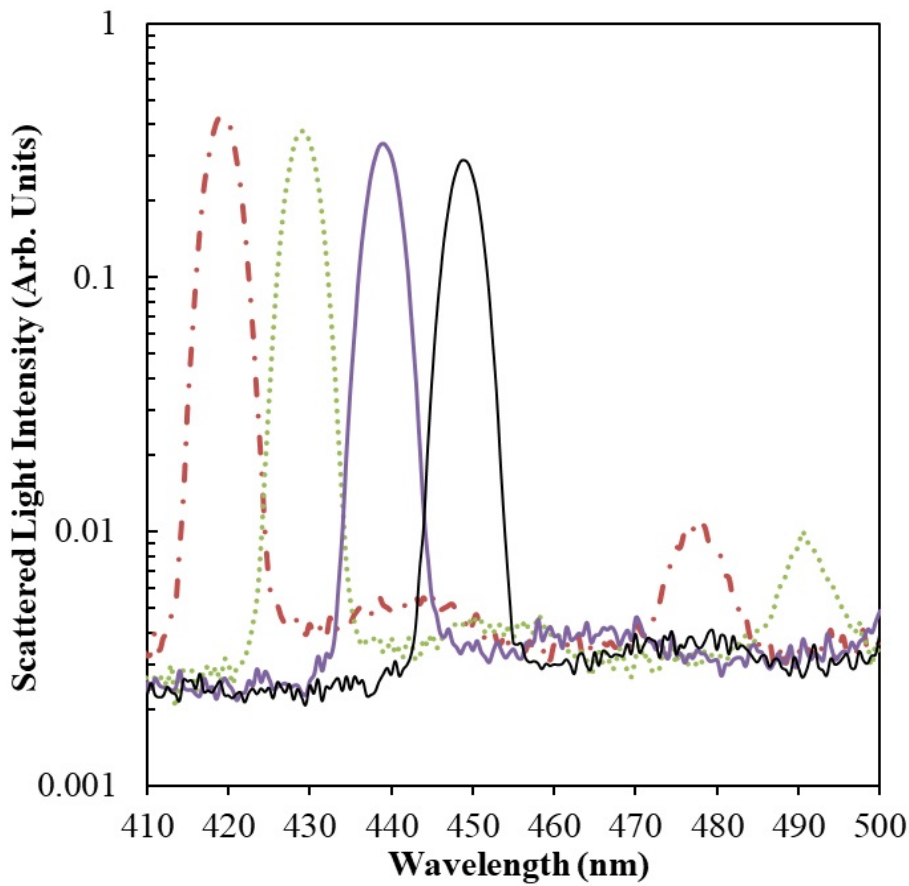

### 3.2 Depolarization Ratio

In the case of fluids composed of isotropic particles, the light scattered in the plane transverse to the plane containing the incident, and scattered beams is completely polarized. However, for anisotropic particles the effect of different polarizability along all the different axes results in an overall increase in the magnitude of scattering [26]. The value of the depolarization ratio for LAB and water have been previously published at certain wavelengths only [27,28,29]. The depolarization ratio for EJ309-base has not been measured or reported.

As a fundamental physical property, the depolarization ratio is defined as the ratio of horizontally to vertically polarized light scattered at 90°. For this study, the depolarization ratio for liquids was measured over a range of wavelengths by measuring the intensities of the polarized light scattering off the sample liquids. The QuantaMaster spectrofluorometer was reconfigured for polarized light measurements. In this configuration, the spectrofluorometer employed Glan – Thompson polarizers in both the excitation and emission optical paths. The excitation polarizer was set to illuminate the sample liquid with vertically polarized light to measure the intensity of vertical (i.e. parallel) component, and the angle of the emission polarizer was then alternately set to measure the intensity of the horizontal (i.e. perpendicular) component of the light scattering off the sample. The ratio of perpendicular to the parallel component of the polarized scattered light at a specific wavelength gives the depolarization ratio at the wavelength. Figure 5. shows the measured values of the depolarization ratio for the three pure liquids.

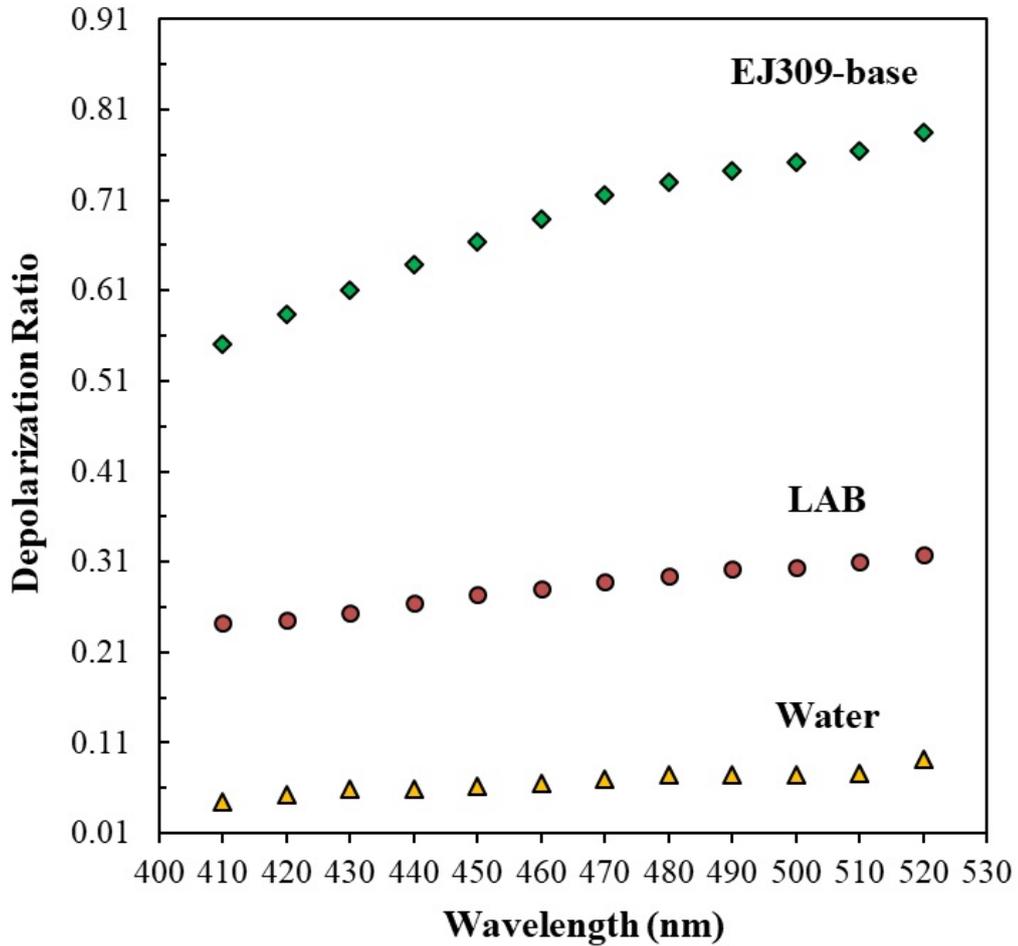

**Figure 5.** Depolarization ratios over a range of wavelengths for the three pure liquid samples. (data in the literatures are only available at λ=430nm: 0.29±0.03, 0.31±0.03, 0.33±0.03 for 3 different types of LABs

### 3.3 Refractive Index

To accurately model the propagation and scattering of light in a liquid based on the *Einstein-Smoluchowski theory*, it is also imperative to know the refractive index, in addition to the depolarization ratio, of the liquid. The index of refraction for a material is dependent on the wavelength of light and the material property of the medium. The variance in the refractive index is known as dispersion, and it is described by the *Sellmeier* formula

$$n^2(\lambda) = 1 + \sum_{i=1}^{N} \frac{B_i}{1 - (C_i/\lambda)^2} \quad (3.3.1)$$

where B and C are *Sellmeier* coefficients from data fitting. For the wavelength range between 410–510 nm, the refractive index of the liquids studied here decreases monotonically with

– 8 –

wavelength λ. The refractive index for both pure water and LAB has been previously determined for a number of different wavelengths [30,31]. It has been shown that for pure organic liquids, the contribution of the temperature fluctuations in the refractive index is negligible [32].

The refractive index at seven different wavelengths was measured for the three pure liquids using a refractometer (ATR – L refractometer manufactured by Schmidt + Haensch GmbH). The *Sellmeier* formula was then fit to these measured refractive index data from 390 to 720 nm. Figure 6. shows that the measured refractive index values (with *Sellmeier* fit) are in good agreement with the published reference values for water.

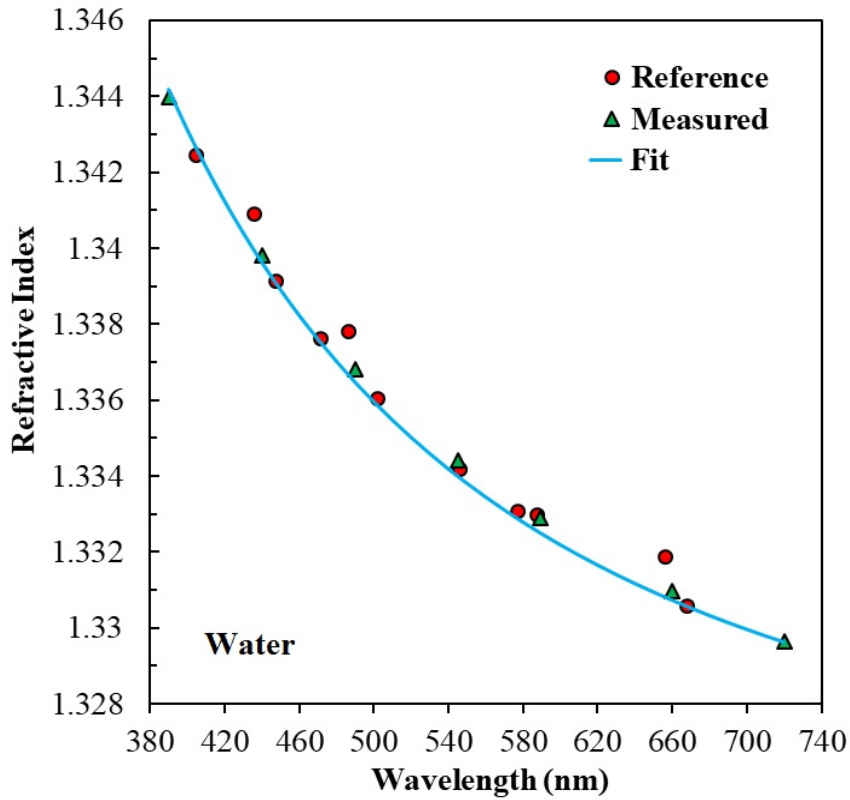

**Figure 6.** Refractive index of pure water. The plot compares the measured data (triangles) and the *Sellmeier* fit (solid line) compared to the reference data (circles).

The *Sellmeier* coefficients were determined using a best-fit model ($R^2 \approx 1$) from the measured refractive index data and then applied to Equation 3.3.1 to describe the refractive index in the wavelength range of 410 – 520 nm. The values of the refractive index for LAB and EJ309-base as a function of wavelength along with the fitted *Sellmeier* formula, are shown in Figure 7.

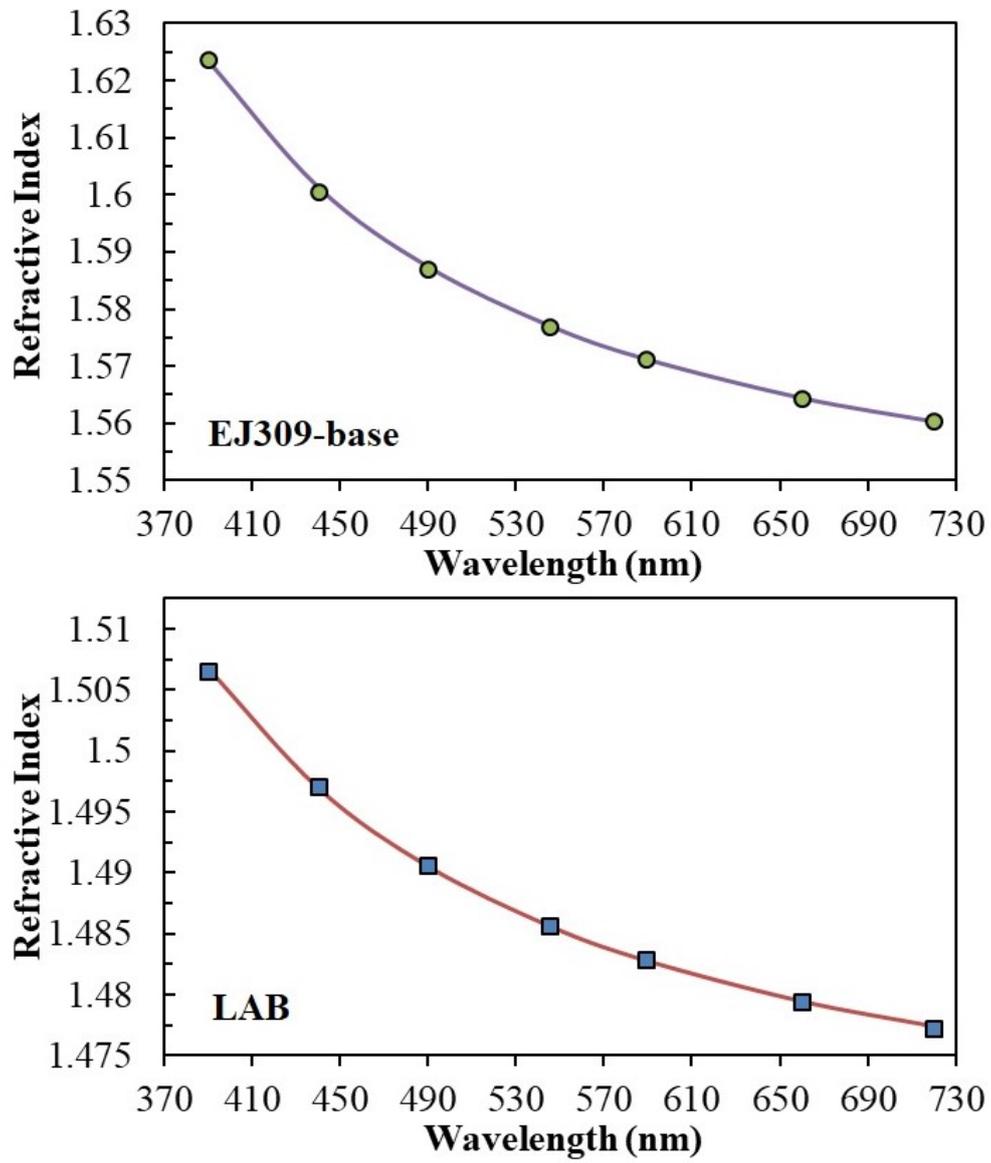

**Figure 7.** Refractive index data for LAB and EJ309-base fit with the *Sellmeier* formula

## 4. Results and Analysis

The scattering length of pure liquid can be determined for a range of wavelengths if the Rayleigh ratios have been determined. In the case of organic liquid scintillators, the measured Rayleigh ratios are dependent on the experimental configuration. Some of these parameters are constants while the rest must be obtained experimentally.

By combining and rearranging equations 2.1, 2.3, and 2.7, we formulated an expression for the intensity of the scattered light, *I*, in equation 4.1. The term *Fitting Constant* is a product of two characteristic parameters: material constant ($C_{Mater}$) and geometry constant ($C_{Geom}$).

$$I = Fitting\ Constant \cdot \left\{ \frac{\pi^2}{\lambda^4} \left[ \frac{(n^2-1)(2n^2+0.8n)}{n^2+0.8n+1} \right]^2 \frac{(6+6\delta)}{(6-7\delta)} \right\} \quad (4.1)$$

The term $C_{Mater}$ accounts for the effect that the material properties have on the intensity of the Rayleigh scattered light. The scattered length can then be computed from $C_{Mater}$, depolarization ratio ($\delta$), and refractive index (*n*) as

$$l_{Ray} = \frac{6}{8\pi} \cdot \frac{1}{C_{Mater}} \cdot \frac{(1+\delta)}{(2+\delta)} \cdot \left\{ \frac{\pi^2}{\lambda^4} \left[ \frac{(n^2-1)(2n^2+0.8n)}{n^2+0.8n+1} \right]^2 \frac{(6+6\delta)}{(6-7\delta)} \right\}^{-1} \quad (4.2)$$

The value of the material constant for water $C_{Mater}^{Water}$ was obtained based on previously published scattering length data and the presently measured values of *δ* and *n*. There are several studies, which have reported different sets of data for the attenuation length of pure water. Most of the works report data measured on laboratory-scale experimental geometries. The wavelength-dependent Rayleigh scattering length data of pure water employed in the Super-Kamiokande experiment on the other hand were measured in an actual detector geometry, and as such it is interpreted as the most accurate measurement of the scattering length for pure water. Therefore, the Super-Kamiokande data set was used as the reference values for this work [34, 35, 36].

Figure 8 shows the measured and fitted Rayleigh scatter intensity data for pure water. Each data point represents the value of the peak height taken from the scattered light spectrum at each wavelength of incident light. The best fit ($R^2$= 0.9955) performed on this data using equation 4.1, yields a value for the *Fitting Constant*, which can then be used to decouple the constants related to the system parameters, $C_{Geom}$.

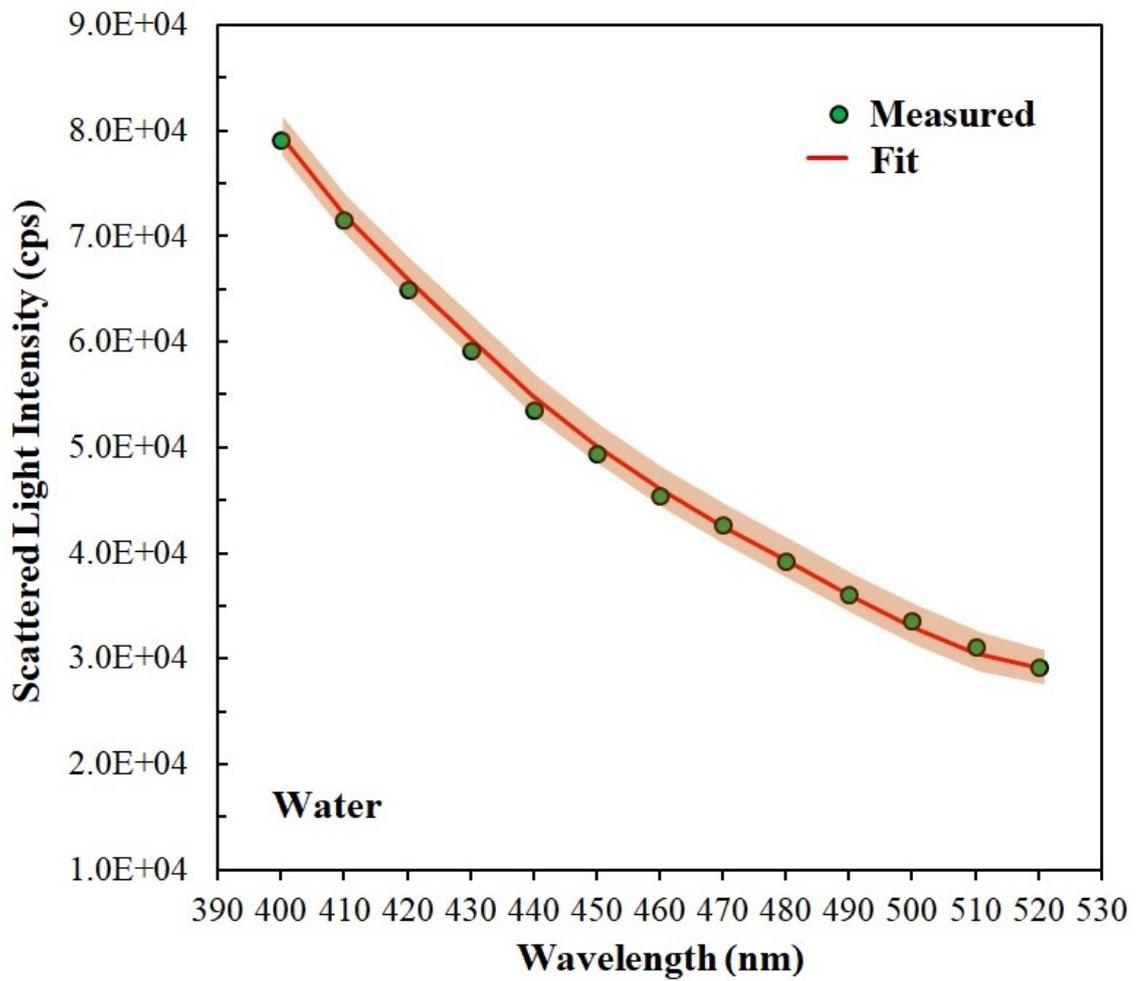

**Figure 8.** Plot showing the peak height intensities of scattered light measured for pure water for different incident light wavelengths and the best fit line obtained using equation 4.1 for the data in the wavelength range of 370 – 510 nm. The goodness of fit was verified by calculating the $R^2$ parameter ($R^2 = 0.9955$).

The best fit ($R^2$=0.9889) of the measured Rayleigh scattered intensity data for LAB was obtained by determining the value for $C_{Mater}^{LAB}$, while the system parameter ($C_{Geom}$) decoupled previously was held as the constant. The derived $C_{Mater}^{LAB}$ and the measured values of $\delta_{LAB}$ and $n_{LAB}$ were then used to determine the Rayleigh scattering length of LAB at each wavelength of interest using equation 4.2. Figure 9 shows the measured scattered light intensity and the corresponding fitted Rayleigh scattering equation for LAB. For incident light up to 440 nm wavelength, the scattered light intensity spectra show the influence of light emission from the excitation of the LAB molecules (see Figure 4), which appears as a divergence of the measured intensity from the $\lambda^{-4}$ dependence. The data fitting was therefore performed for wavelengths >440 nm and then extrapolated back to the range of 410–440 nm.

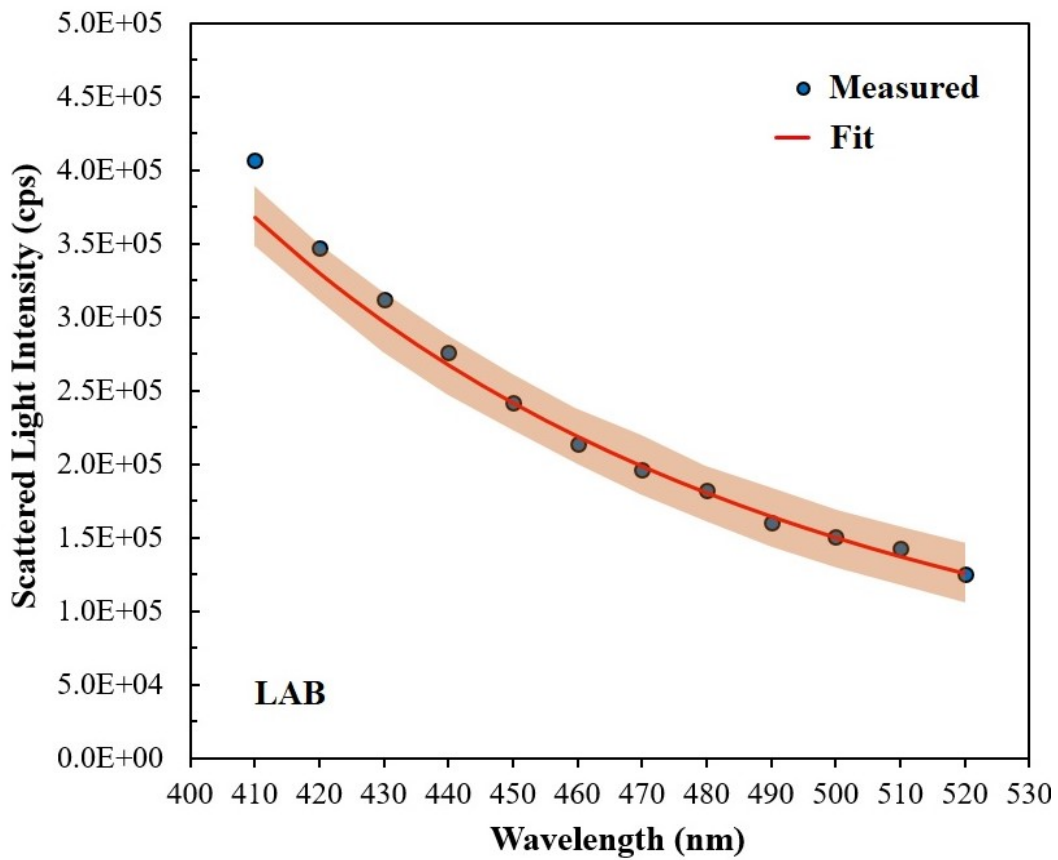

**Figure 9.** Measured scattered light intensity for LAB and the best fit using equation 4.1 for the Rayleigh scattered intensity in the wavelength range of 450 – 520 nm. The goodness of fit was verified by calculating the $R^2$ parameter ($R^2 = 0.9889$). The 95% confidence interval was computed using the standard error of the fit and is depicted as the band along the fitted curve.

EJ309-base has greater light scattering intensity compared to LAB, as also seen from greater anisotropy depolarization ratio than that of LAB (see Figure 5), and therefore its scattering length will presumably be shorter. Unlike the LAB, the EJ309-base had not been subjected to additional purification or filtering steps for this present work. Figure 10 shows the measured scattered light intensities for EJ309-base and the corresponding fitted ($R^2=0.9875$) Rayleigh scattering equation. As in the case of LAB shown in Figure 4, the Rayleigh scattering intensity formula was fit to only the measured data above the wavelength of 440 nm to avoid the influence of emitted scintillation light on the scattered light intensity.

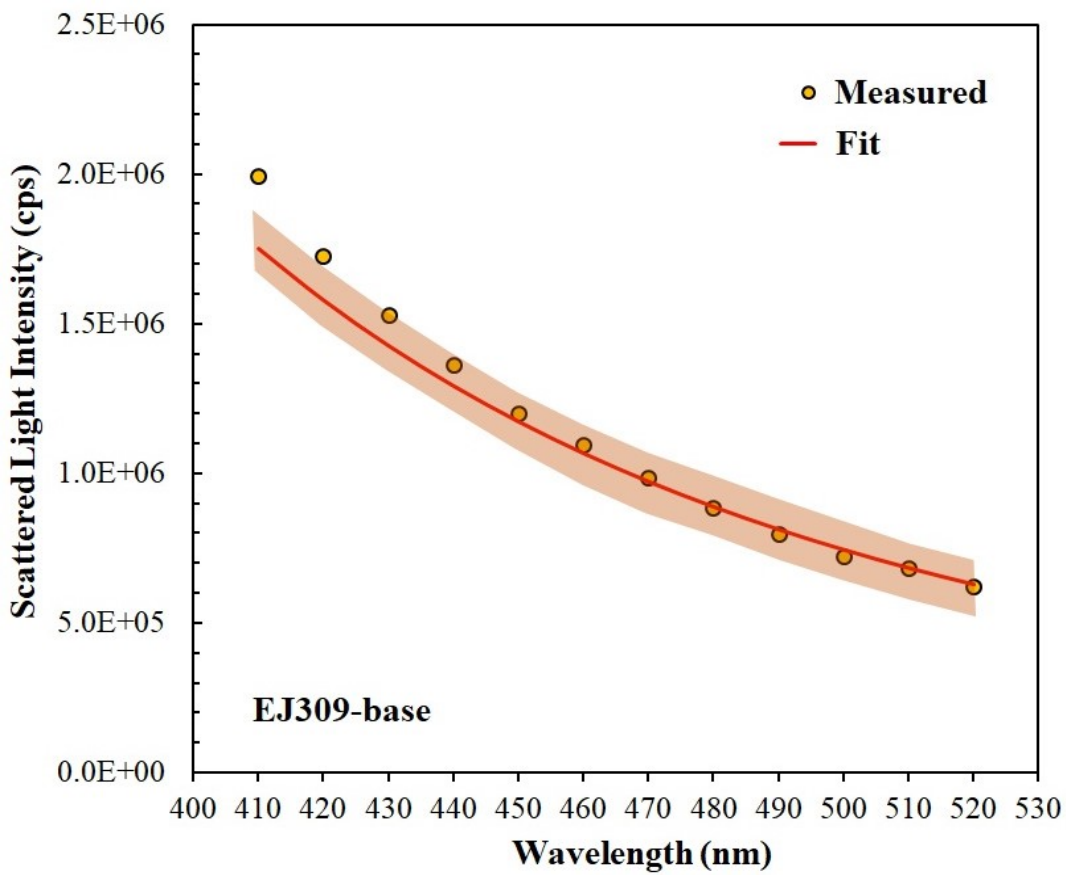

**Figure 10.** Measured scattered light intensity and for EJ309-base and the best fit using equation 4.1 for the Rayleigh scattered intensity in the wavelength range of 450 – 520 nm. The goodness of fit was verified by calculating the $R^2$ parameter ($R^2 = 0.9875$). The 95% confidence interval was computed using the standard error of the fit and is depicted as the band along the fitted curve.

The propagation of uncertainty for scattering length is the effect of variables′ uncertainties from the scattered light intensity, the depolarization ratio, and the refractive index measurements. The scatter in the measured values of refractive index is a small fluctuation in the order of ±0.0006% of the mean value, therefore its contributed error to the scattering length uncertainty is negligible.

Extensive use was made of the automation capabilities of the QuantaMaster-8000 Fluorescence Spectrometer. The wavelength accuracy of this spectrofluorometer is ±0.3 nm, as described in [25]. The geometry of the optical components and slit width to control the light alignment and intensity were kept constant to minimize systematic fluctuations for every set of measurement. While measuring the scattered light intensity and the depolarization ratio, the cuvette was rinsed and dried before being used for each sample. During this procedure, the position alignment of the cuvette was also fixed by the holder in the sample chamber.

For a typical set of depolarization ratio or scattering light-intensity measurement, three samples were prepared and assayed. Accordingly, the mean value and its corresponding standard deviation, related to the overall (optical or geometric) variables in the system, at the wavelength of interest were obtained. The uncertainty of scattering length is then calculated as

$$\sigma_{l_{Rayleigh}} = \sqrt{\left(\sigma_{Scattered-Light-Intensity}\right)^2 + \left(\sigma_{Depolarization-Ratio}\right)^2}$$

The Rayleigh scattering lengths of LAB and EJ309-base in the wavelength region of 410 – 520 nm computed using Equation 4.2 are reported in Table 1.

**Table 1. Rayleigh scattering length as a function of wavelength for LAB and EJ309-base.**

| Wavelength (nm) | LAB (m) | EJ309-base (m) |
|:---:|:---:|:---:|
| 410[+] | 22.9 ± 2.3 | 5.0 ± 0.5 |
| 420[+] | 25.3 ± 2.5 | 5.5 ± 0.5 |
| 430[+,x] | 27.9 ± 2.3 | 6.1 ± 0.6 |
| 440[+] | 30.6 ± 2.6 | 6.7 ± 0.5 |
| 450 | 33.6 ± 3.1 | 7.4 ± 0.7 |
| 460 | 36.8 ± 3.6 | 8.1 ± 0.8 |
| 470 | 40.2 ± 3.3 | 8.9 ± 0.7 |
| 480 | 43.9 ± 4.1 | 9.7 ± 1.0 |
| 490 | 47.8 ± 5.5 | 10.6 ± 1.2 |
| 500 | 51.8 ± 5.8 | 11.5 ± 1.2 |
| 510 | 56.2 ± 7.1 | 12.5 ± 1.3 |
| 520 | 60.9 ± 7.7 | 13.6 ± 1.6 |

[+]Scattering lengths extrapolated from the fitted formula using data from wavelengths of >440nm
[x]Scattering lengths for LAB available in the literatures: 26.2 ± 1.9 [27] and 28.5 ± 2.3 [28]

## 5. Discussion and Conclusions

The Rayleigh scattering length of a pure liquid can be determined using an approach derived from the *Einstein-Smoluchowski* theory. Several optical parameters: the depolarization ratio, the refractive index, and the Rayleigh scattered light intensity, were experimentally measured over a range of incident light wavelengths.

This study reported scattering lengths for LAB and EJ309-base in the wavelength-emitted region from 410 to 520nm of an nLS. The data presented here for LAB are also in good agreement with previously published data although those are only available at few wavelengths [27,28]. In the case of EJ309-base, this work is the first publication of wavelength-dependent scattering lengths for a DIN-based LS. The $\lambda^{-4}$ dependence of Rayleigh scattering was clearly demonstrated by the measurements even for pure water; however, some effects from the absorption-reemission process were also observed for the organic liquid scintillator samples below 440nm.

Scintillation and optical transport in a large-volume liquid scintillator detector involve a complex interdependence of many parameters, and yet the optical property, such as the scattering length, allows some broad conclusions to be made regarding the performance of liquid scintillator. The measured Rayleigh scattered length is an imperative input parameter for optical transport simulations. Based on the results of this study, the scattering length (e.g., at $\lambda$ = 430 nm) of 27.9 ± 2.3 m for LAB confirms its feasibility for use in large-volume detectors, while the shorter scattering length of 6.1 ± 0.6 m for EJ309-base makes it a good scintillator candidate for smaller scale detectors.

**Acknowledgments**

The work conducted at Brookhaven National Laboratory was supported by the U.S. Department of Energy under Contract No. DE-SC0012704. The material was based upon work supported by the U.S. Department of Energy, National Nuclear Security Administration, Office of Defense Nuclear Nonproliferation Research and Development.